\documentclass[A4]{aa}
\usepackage{graphicx}
\usepackage{txfonts}
\begin{document}
\title{First {\it XMM-Newton} study of two Narrow--Line Seyfert 1 galaxies
discovered in the Sloan Digital Sky Survey\thanks{Based on public observations obtained with 
\emph{XMM-Newton}, an ESA science mission with instruments and contributions 
directly funded by ESA Member States and the USA (NASA).}}

\author{L. Foschini\inst{1}, V. Braito\inst{2}, G.G.C. Palumbo\inst{3}, G. Ponti\inst{1,3},\\ 
M. Dadina\inst{1}, R. Della Ceca\inst{2}, G. Di Cocco\inst{2}, P. Grandi\inst{1}, G. Malaguti\inst{1}}

\offprints{L. Foschini \email{foschini@bo.iasf.cnr.it}}

\institute{Istituto di Astrofisica Spaziale e Fisica Cosmica (IASF) del
CNR, Sezione di Bologna, Via Gobetti 101, 40129 Bologna (Italy)
\and
INAF -- Osservatorio Astronomico di Brera, Via Brera 28, 20121 Milano (Italy)
\and
Dipartimento di Astronomia, Universit\`a di Bologna, via Ranzani 1, 40127 Bologna (Italy)
}

\date{Received 02 July 2004; accepted 01 August 2004}

\abstract{The Early Data Release of the Sloan Digital Sky Survey (SDSS) contains 150 Narrow--Line 
Seyfert 1 (NLS1) galaxies, most of them previously unknown. We present here the study of the 
X--ray emission from two of these active galaxies (SDSS J$030639.57+000343.2$ and 
SDSS J$141519.50-003021.6$), based upon \emph{XMM-Newton} observations. The spectral and timing 
characteristics of the two sources are presented and compared against the typical properties
of known NLS1 galaxies. We found that these two NLS1 are within the dispersion range of
the typical values of this class of AGN, although with some interesting features that deserve
further studies.
}

\authorrunning{L. Foschini et al.}
\titlerunning{\emph{XMM-Newton} observations of two Narrow--Line Seyfert 1...}
\maketitle

\section{Introduction}
The first studies of Markarian 359 (Davidson \& Kinman 1978) and of similar objects led to the 
identification of a peculiar subclass of Seyfert galaxies: the Narrow-Line Seyfert 1 
galaxies (NLS1) (Osterbrock \& Pogge 1985, Goodrich 1989). Such objects presented an 
intriguing mixture of physical phenomena, which made them very attractive and popular during the 
last decade. Their distinctive optical characteristics are $FWHM(\rm H\beta) < 2000$ km/s, 
weak [OIII] and strong Fe II relative to H$\beta$, particularly [OIII]/H$\beta <3$ (see Grupe 2000, and 
Pogge 2000 for a review). 

Stephens (1989) suggested that X--ray selection could be an efficient method to identify NLS1. 
Since then, many efforts have been devoted in analysing X--ray data of NLS1, mainly thanks to the 
satellites ROSAT, ASCA, and more recently \emph{XMM-Newton} and \emph{Chandra}. NLS1 typical characteristics in the X--ray domain are the presence 
of a soft X-ray excess, probably the high-energy tail of the radiation emitted from the accretion disk, 
a pronounced variability (NLS1 are the most variable AGNs known after blazars), and a steep photon index (e.g. 
Puchnarewicz et al. 1992, Boller et al. 1996, Wang et al. 1996, Grupe et al. 1998, 1999, Leighly 1999a,b).
From all these studies, it resulted that NLS1 could have the smallest central BH mass among all AGN 
(as low as $10^5$ $M_{\odot}$) and super--Eddington accretion rates (e.g. Wang et al. 1996). 

Since most of NLS1 are X-ray selected, it is not clear if the above mentioned characteristics 
are typical of these objects or are the result of a biased selection. For example, it is worth mentioning 
that some NLS1 could have X-ray behaviour similar to Seyfert 1 (see Fig. 8 of Boller et al. 1996). 
Therefore, it is necessary to try the inverse operation of selection, i.e. to study the X-ray characteristics of a large
sample of optically selected NLS1. In this respect, the 150 NLS1 found in the Early Release of the Sloan 
Digital Sky Survey (SDSS\footnote{\texttt{http://www.sdss.org/}}) by Williams et al. (2000) 
represent a valuable starting point. 
  
In order to pursue this goal, the \emph{XMM-Newton} public archive has been searched for coincidental
position with SDSS galaxies of the sample of Williams et al. (2000). Two SDSS sources, SDSS J$141519.50-003021.6$ ($z=0.135$) and 
SDSS J$030639.57+000343.2$ ($z=0.107$), turned out to be in the \emph{XMM-Newton} observed fields. 
The distance of the sources from the boresight was $603''$ and $170''$, respectively.
For the sake of simplicity, they will be thereafter called SD1 and SD2, respectively. 
These sources were already observed in X--rays only by ROSAT (see Williams et al. 2002), and
therefore the analysis of \emph{XMM-Newton} data presented here represent the first
$2-10$ keV study.

In this paper, $H_{0}=70$ km s$^{-1}$ Mpc$^{-1}$ is assumed throughout.

\begin{figure*}[!ht]
\centering
\includegraphics[scale=0.35,angle=270]{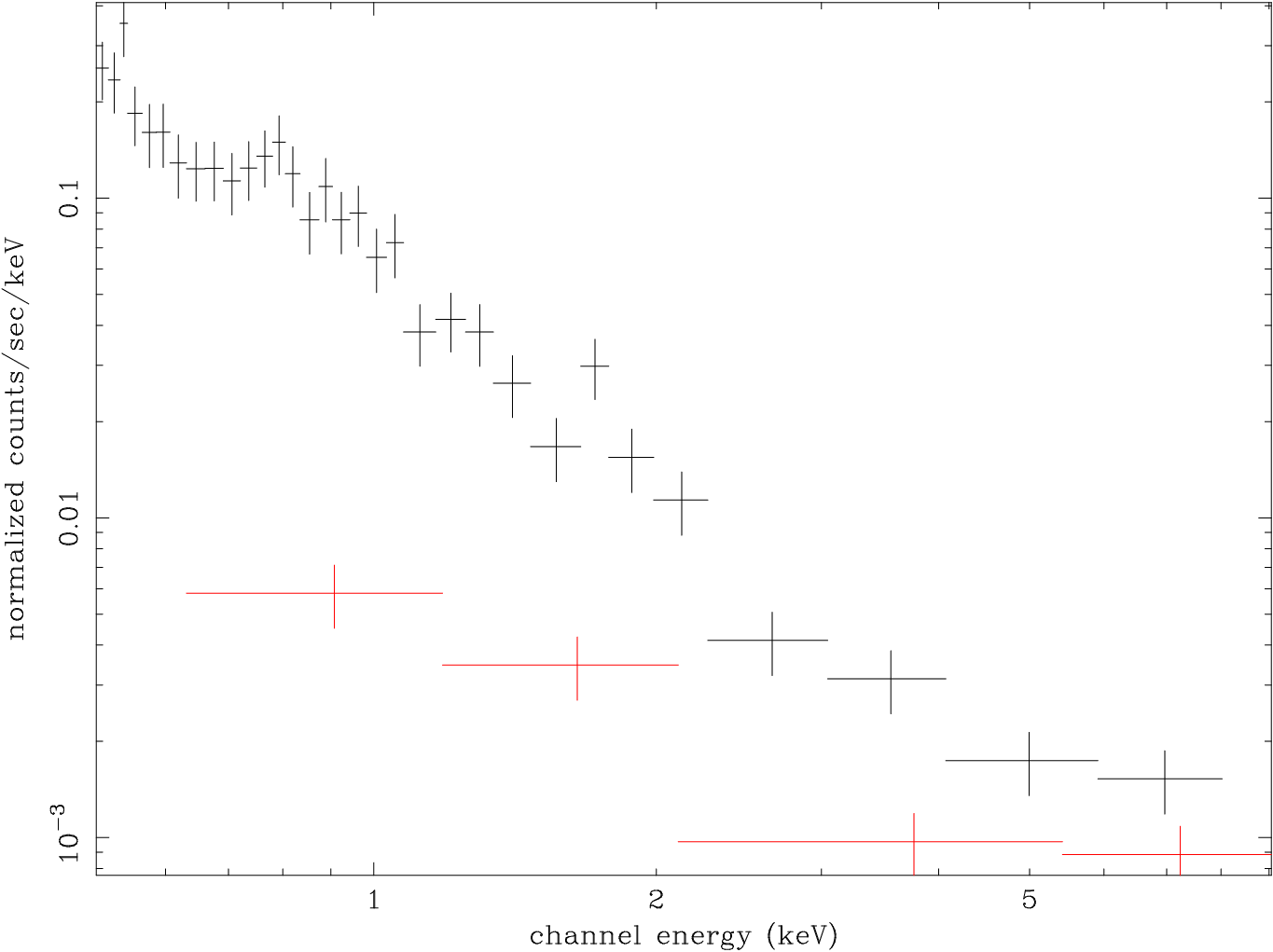}
\includegraphics[scale=0.35,angle=270]{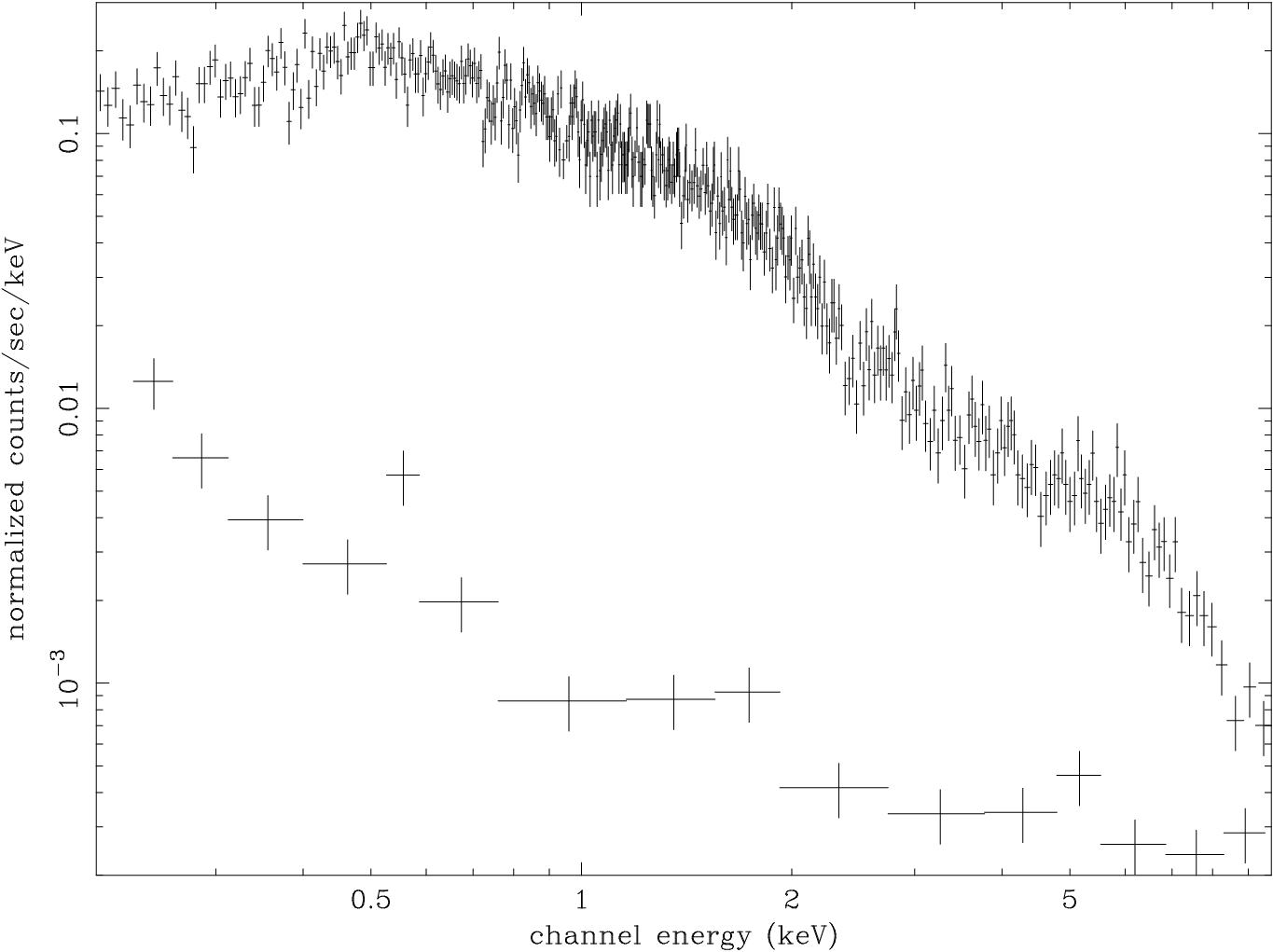}
\caption{EPIC--PN comparison of source and background spectra for SD1 (\emph{left}) 
and SD2 (\emph{right}). The upper data refer to the source
and background, while the lower one is the background only extracted from a region equal to that
of the source.}
\label{compsd2}
\end{figure*}

\section{$XMM-Newton$ observation and data reduction}
For the processing, screening, and analysis of the data from the EPIC MOS1 and
MOS2 (Turner et al. 2001) and PN cameras (Str\"uder et al. 2001),
standard tools have been used (\texttt{XMM SAS v. 5.4.1} and
HEAsoft \texttt{Xspec 11.3.1} and \texttt{Xronos 5.19}) and 
standard procedures described in Snowden et al. (2002) followed. The 
observations of both sources were affected by solar soft-proton flares, so that a  
cleaning was necessary.

To study variability, the lightcurves from EPIC-PN data were extracted, since this
detector has the best time resolution in full frame mode ($\sim 73$~ms) and the highest
throughput. The source region was centered in the optical position of the two NLS1, $\alpha=14:15:19.50$, 
$\delta=-00:30:21.6$ for SD1, and $\alpha=03:06:39.57$, $\delta=+00:03:43.2$ for SD2 
(J2000, SDSS uncertainty $0.1''$, Pier et al. 2003), with a radius of $40''$ and $35''$, respectively. 
The background to be subtracted in the analysis was derived from an annular region with maximum 
radius of $2'$ for SD1, and from a circular region $1'$ wide near the source for SD2. The different 
radii of extraction were due to the position of the source in the detector chip (closeness to gaps).

Data from the same regions were also used for the spectral analysis.
The spectra were rebinned so that each energy bin contained a minimum of 25 counts, and
fitted only in the $0.5-10$~keV energy range because of the uncertainties in the MOS cameras calibration
at lower energies (cf. Kirsch 2003). The photon redistribution matrix and the related ancillary
file were created appropriately with the \texttt{rmfgen} and \texttt{arfgen} tasks of XMM-SAS.

Despite the high background, and the low statistics (particularly for SD1), the signal 
is significantly higher when compared to the background level, up to high energy 
(Fig.~\ref{compsd2}). Therefore, any feature at high energy can be due to the source.

\begin{figure}[!ht]
\centering
\includegraphics[scale=0.35,angle=270]{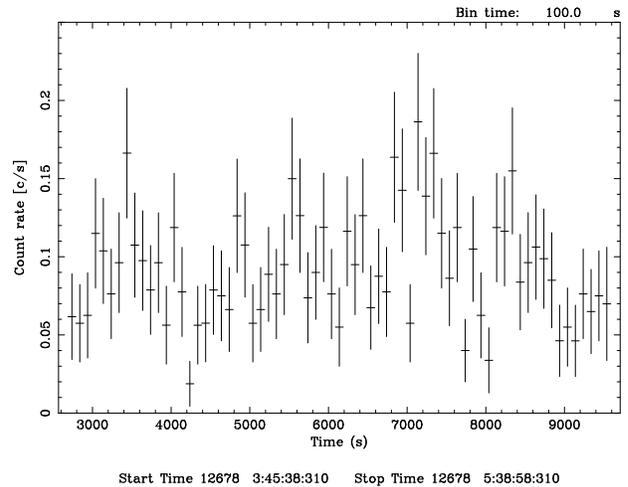}
\caption{EPIC--PN global light curve of SD1, after cleaning for soft-proton flares, binned at $100$~s. 
Error bars are at $1\sigma$.}
\label{lcsd1}
\end{figure}

\begin{figure*}[!ht]
\centering
\includegraphics[scale=0.35,angle=270]{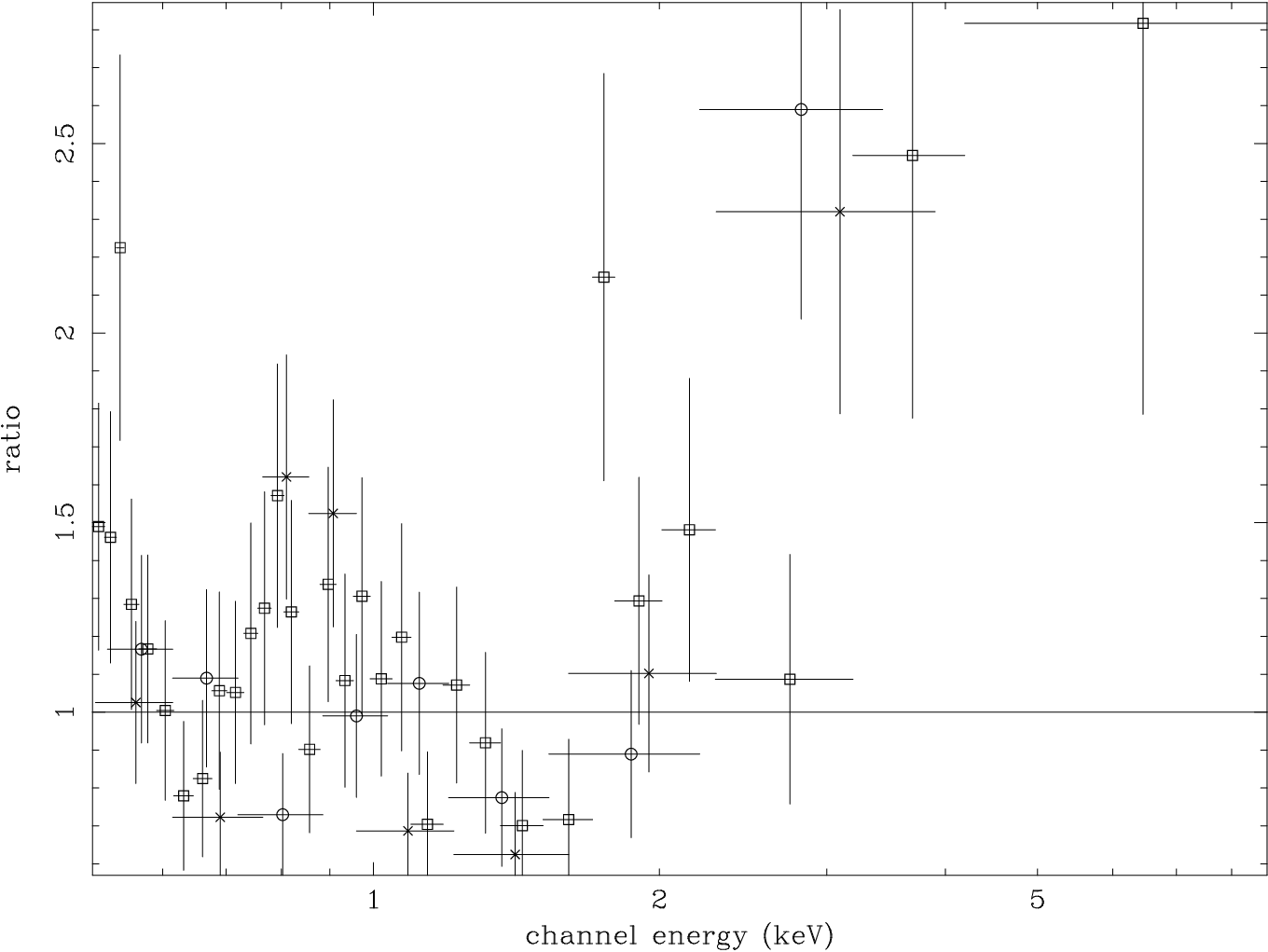}
\includegraphics[scale=0.35,angle=270]{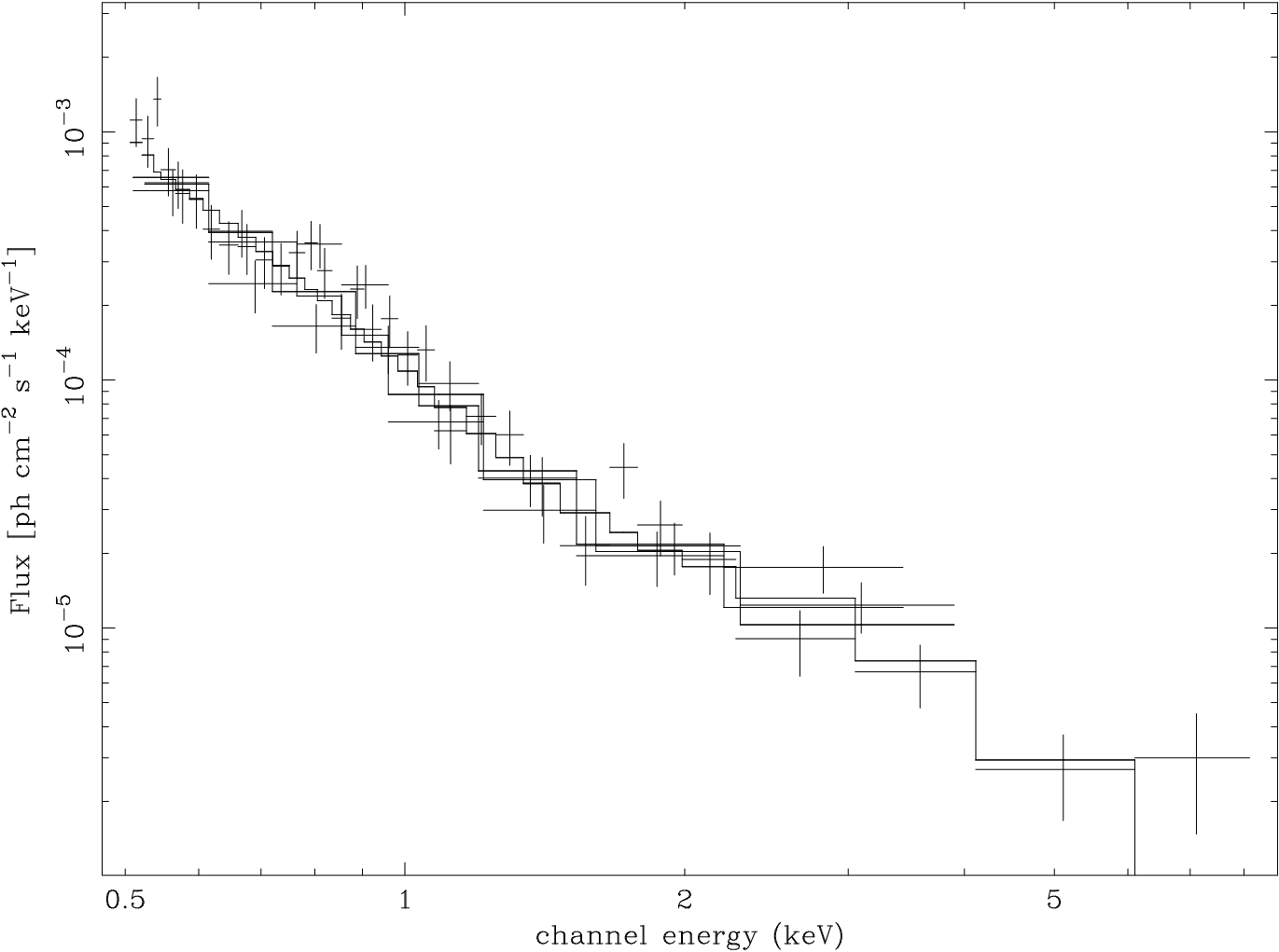}
\caption{(\emph{left}) Ratio data/model of the EPIC data of SD1 fitted with only a power 
law absorbed by the Galactic 
column. Circles are MOS1 data, crosses are MOS2 data, and squares are PN data. 
The soft and hard excesses are clearly visible. Data are rebinned for clarity. (\emph{right})
Unfolded spectrum of SD1 with the best fit model (partial covering).}
\label{sd1}
\end{figure*}

\section{SDSS J$141519.50-003021.6$ (SD1)}
The source SD1 was in the field of view of the ObsID $0145480101$ during the observation 
performed on $2003-08-02$. All the EPIC detectors were set in full frame mode and
collected data for an elapsed time of $23567$ s. However, the presence of soft--proton
flares reduced the effective exposure to $6.7$ ks for both MOS and $6.1$ ks for PN.

The source displays a hint of variability (Fig. \ref{lcsd1}), with the $\chi^2$ probability of 
constancy less than $7$\%, and of a $476$~s periodicity ($2\sigma$) in the energy band $0.5-2$~keV. 
However, the low statistics prevents more detailed studies.
Longer and unflared observations are needed to assess this feature.

We started the spectral analysis by fitting the average spectrum integrated over the whole
effective exposure, with a simple power law model and the Galactic absorption
along the line of sight ($N_{\rm H}=3.2\times 10^{20}$~cm$^{-2}$, Dickey \& Lockman 1990). 
This resulted in a steep ($\Gamma=2.9\pm 0.2$) spectrum, but with a low statistical quality 
($\chi^2=81.4$, dof$=46$). The ratio 
data/model clearly shows excesses below $0.6$~keV and above $2$~keV (Fig.~\ref{sd1}). 
The excess at high energy cannot be due to residual background, as shown in Fig.~\ref{compsd2}.

The addition of a redshifted thermal component (\texttt{zbb} model in \texttt{xspec}) with $kT=0.14\pm 0.02$~keV 
improved significantly the fit ($\chi^2=54.6$, dof$=44$). In this case, the photon index
of the power law model decreases to $\Gamma=1.8\pm 0.2$, but still with some residuals
at high energy. The fit with a gaussian emission line fixed at $E=6.5$ keV with 
a width $\sigma=1$ keV was rejected ($\chi^2=54.8$, dof$=43$), and also other models, like
the constant density ionized model by Ross \& Fabian (1993), or the reflection by
ionized material by Magdziarz \& Zdziarski (1995), were not successfully constrained.

A partial covering (\texttt{pcfabs} model in \texttt{xspec}) appears to be the best fit model 
($\chi^2=51.0$, dof$=44$), with $N_{\rm H}=(4_{-1}^{+3})\times 
10^{22}$~cm$^{-2}$, a covering factor of $0.84\pm 0.01$, and a photon index $\Gamma=3.4\pm 0.2$.
In this case, the observed flux in the energy band $0.5-10$ keV is 
$4.6\times 10^{-13}$~erg~cm$^{-2}$~s$^{-1}$, corresponding to an intrinsic luminosity 
of $1.2\times 10^{44}$~erg/s.

\section{SDSS J$030639.57+000343.2$ (SD2)}
The source SD2 was observed on $2003-02-11$ in the field of view of the ObsID $0142610101$,
with an elapsed time of $73918$ s. All the EPIC camera detectors were set in full frame mode.
Because of the presence of soft--proton flares, the effective exposures were about $52$ ks for both MOS
and $38$ ks for PN. 

\begin{figure}[!ht]
\centering
\includegraphics[scale=0.35,angle=270]{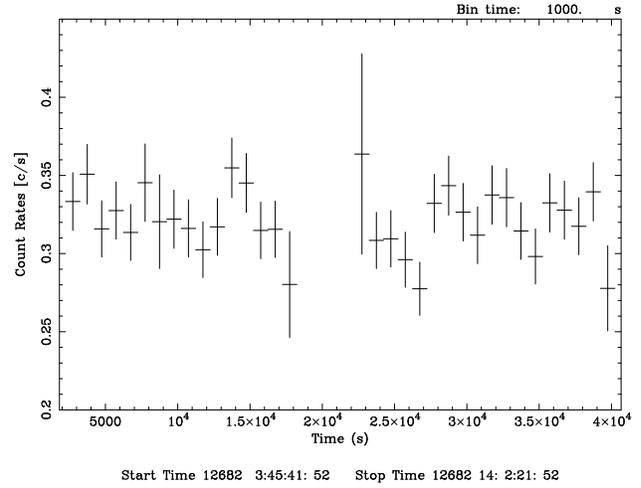}
\caption{EPIC-PN global light curve of SD2, after cleaning for soft-proton flares, binned at $1000$~s. 
Error bars are at $1\sigma$.}
\label{lcsd2}
\end{figure}

The source does not show statistically significant variability (Fig.~\ref{lcsd2}), the $\chi^2$ 
probability of constancy being greater than $92$\%, depending on the bin width ($100-1000$~s). 
This appears to be unusual, given the well known variability 
of the NLS1 in the X-ray energy band: Leighly (1999a) reported variability with at least $99$\%
confidence level.
The steady flux is, however, understandable if the cause of variability is
mainly a flare--like phenomenon. In this case, we were observing a period without flares.

\begin{figure*}[!ht]
\centering
\includegraphics[scale=0.35,angle=270]{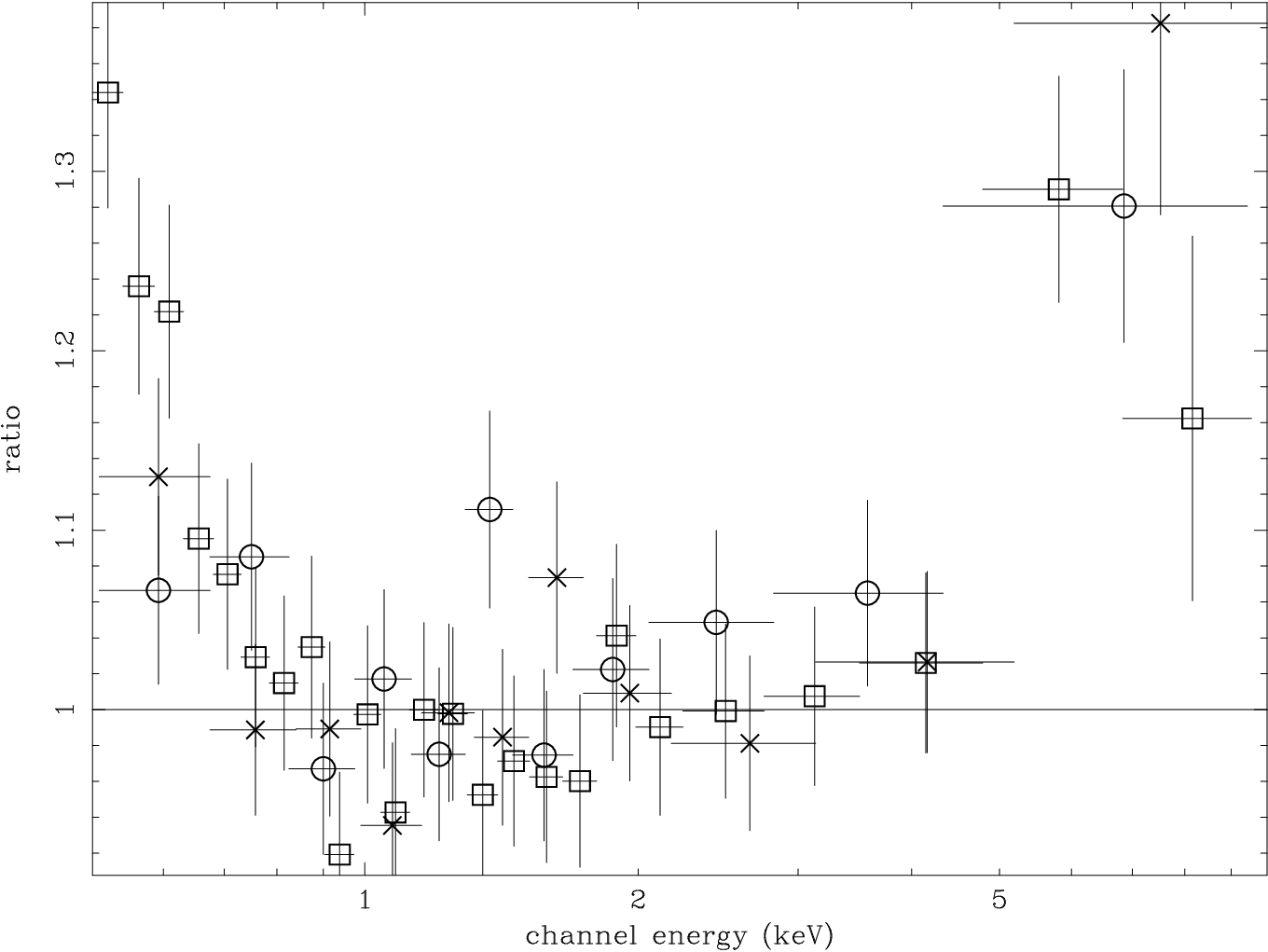}
\includegraphics[scale=0.35,angle=270]{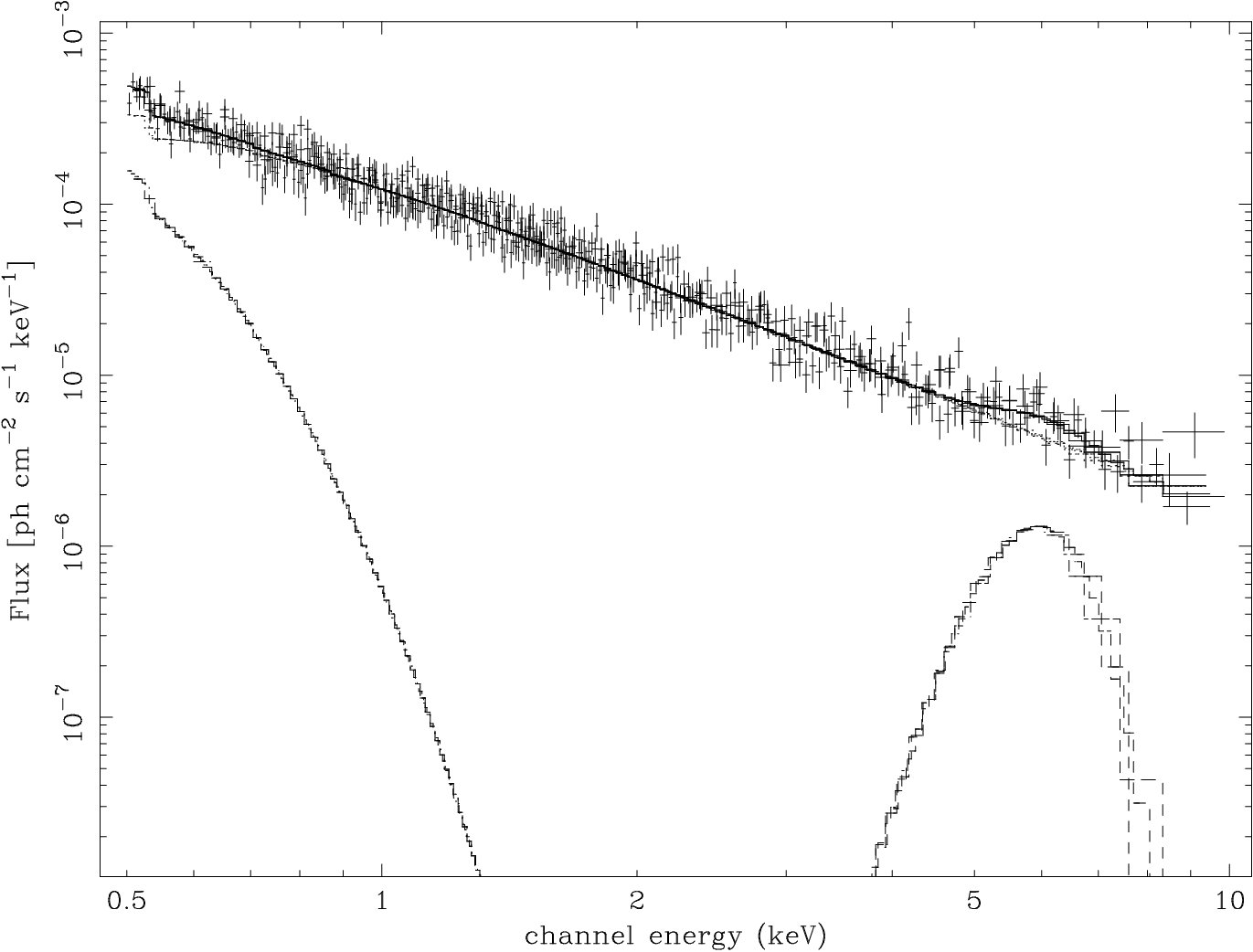}
\caption{(\emph{left}) Ratio data/model of the EPIC data of SD2 fitted with only a power law 
absorbed by the Galactic column. Circles are MOS1 data, crosses are MOS2 data, and squares are PN data. 
The soft and hard excesses are clearly visible. Data are rebinned for clarity. (\emph{right})
Unfolded spectrum of SD2 with the best fit model (power law plus black body and gaussian emission line).}
\label{compsd3}
\end{figure*}

Therefore, it is reasonable to study only the averaged spectrum integrated over the whole
period. 
The first fit was obtained with a simple power law model ($\Gamma=1.99\pm 0.02$), with the Galactic 
absorption toward SD2 ($N_H=6.94\times 10^{20}$ cm$^{-2}$, Dickey \& Lockman 1990). The result 
is shown in Fig. \ref{compsd3}: soft and hard excesses are clearly visible ($\chi^2=806.9$, 
dof$=561$). To fit the soft excess we adotped the redshifted black body (\texttt{zbb} model
in \texttt{xspec}) with $kT=0.083\pm 0.004$~keV. The photon index decreases to $\Gamma=1.91\pm 0.03$,
and the fit improves to $\chi^2=739.9$ for dof$=559$. The thermal component is required
at $>99.99$\%. 

An excess at energies greater than $\sim 5$~keV remained in the data still after
having used different procedures in the cleaning for soft--proton flares, but the comparison 
with the background (Fig.~\ref{compsd2}) confirms that the excess is real. 
To model this hard X-ray excess, we tried with a simple large
gaussian line at $E=6.6_{-0.5}^{+0.2}$~keV, $\sigma=0.8_{-0.3}^{+0.9}$~keV, and
equivalent width $510_{-234}^{+1030}$~eV. This line is required at $99.15$\% level, and
improves the fit to $\chi^2=724.5$ for $556$ degrees of freedom. The thermal component is now 
$kT=0.079_{-0.005}^{+0.004}$~keV and the photon index is $\Gamma=1.95_{-0.04}^{+0.05}$. 
This is the best fit model for this source (Fig.~\ref{compsd3}, \emph{right}). The observed flux in the energy band 
$0.5-10$~keV is $6.9\times 10^{-13}$~erg~cm$^{-2}$~s$^{-1}$, corresponding to an intrinsic 
luminosity of $2.3\times 10^{43}$~erg/s.

We tried also a partial covering, obtaining $N_{\rm H}=(19_{-9}^{+15})\times 10^{22}$~cm$^{-2}$,
a fraction of $0.37_{-0.10}^{+0.09}$, and a photon index $\Gamma=2.07\pm 0.04$. The fit,
with $\chi^2=766.7$ for $559$ degrees of freedom, is worse than the previous one.

Also the Ross \& Fabian (1995) model does not provide a good fit ($\chi^2=749.0$, dof$=559$). 
In this case, we found that the ionization parameter is $\log \xi=3.25_{-0.02}^{+0.15}$,
with a reflection fraction of $0.5_{-0.1}^{+0.2}$, and $\Gamma=1.85\pm 0.04$.

\section{Observations at other wavelengths}

\subsection{SDSS J$141519.50-003021.6$ (SD1)}
The optical counterparts of SD1 have been found in 2MASS (Cutri et al. 2003) and USNO B1 
(Monet et al. 2003) catalogs. The 2MASS position is $RA=14:15:19.51$ and $Dec=-00:30:21.1$
(J2000, uncertainty $2''$), consistent with that of SDSS. The source was detected with
magnitudes $J=16.7\pm 0.2$, $H=16.5\pm 0.2$, and $K=15.4\pm 0.2$. The USNO B1 catalog
provides the magnitudes in two periods: (1) Palomar Observatory Sky Survey (POSS) I, 
($1949-1965$), with emulsion sensitive at wavelengths in the range $620-670$ nm; (2) POSS II 
($1985-2000$), sensitive at $385-540$ nm. These are: $B_1=19.3\pm 0.3$, $B_2=19.6\pm 0.3$,
and $R_1=18.6\pm 0.3$, $R_2=18.6\pm 0.3$, plus $I=17.9\pm 0.3$. The detections in the
two periods do not show variability within the measurement errors. 
Also in this case, the position found ($RA=14:15:19.5$, $Dec=-00:30:21.4$, J2000) 
matches that of the SDSS.

\subsection{SDSS J$030639.57+000343.2$ (SD2)}
For this source, in addition to the counterparts in the 2MASS and USNO B1 catalog,
a radio counterpart was also found in the FIRST\footnote{\texttt{http://sundog.stsci.edu/}}
and NVSS\footnote{\texttt{http://www.cv.nrao.edu/nvss/}} surveys, both at $1.4$ GHz ($20$ cm). The FIRST survey 
was performed between $1993$ and $2002$, while the NVSS was done between $1993$ and $1997$. 
The source appears to be compact, with a peak flux is $4.4\pm 0.4$ mJy (NVSS).

This galaxy appears to be very bright in the optical/infrared wavelengths: the 2MASS
catalog reports a clear detection with all filters: $J=15.08\pm 0.08$, $H=14.12\pm 0.08$,
$K=13.33\pm 0.05$. The USNO B1 catalog gives these values: $B_1=16.5\pm 0.3$, $B_2=14.5\pm 0.3$,
$R_1=14.7\pm 0.3$, $R_2=13.5\pm 0.3$, and $I=14.8\pm 0.3$, showing a substantial degree of variability.
Specifically, it is worth noting the change of about $2$ magnitudes in the B band between the
two reference periods.

\section{Discussion}
We presented here the first analysis in a wide X-ray energy band ($0.5-10$ keV) performed
to date on two Narrow--Line Seyfert 1 Galaxies from the optically selected sample of Williams
et al. (2002). By comparing the spectral parameters obtained for the present sources with
the average values found by Leighly (1999a,b), it is clear that the X-ray characteristics
are in the ranges of the NLS1. The mean value of the photon index found by Leighly (1999b)
is $2.19\pm 0.10$, with a dispersion of $0.30_{-0.06}^{+0.07}$, for NLS1, and
$1.78\pm 0.11$, with a dispersion of $0.29_{-0.07}^{+0.09}$ for Seyfert 1 active nuclei. 
These values should be compared
with the photon index of $3.4$ for SD1 and $1.95$ for SD2. SD1 appears steeper than the mean
value of Leighly, but it is worth noting that a steep photon index can still be likely
(see Fig. 8 by Boller et al. 1996).

Only SD1 displays a certain degree variability, although the limited effective exposure ($6$ ks)
prevents further studies. SD2 does not show variability, but this could be explained
with a period of quiescence. To compare the present fluxes with the ROSAT observations,
we converted, with WebPIMMS\footnote{\texttt{http://heasarc.gsfc.nasa.gov/Tools/w3pimms.html}}, 
the PSPC count rates into the flux in the band $0.5-2.4$ keV, by using the Galactic $N_{\rm H}$ 
and the photon index $\Gamma$ reported by Williams et al. (2002). The calculated observed
flux is $(1.4\pm 0.4)\times 10^{-13}$~erg~cm$^{-2}$~s$^{-1}$ for SD1 and 
$(1.1\pm 0.2)\times 10^{-12}$~erg~cm$^{-2}$~s$^{-1}$ for SD2. The corresponding 
fluxes from the present \emph{XMM-Newton} observations are 
$(3_{-1}^{+2})\times 10^{-13}$~erg~cm$^{-2}$~s$^{-1}$ and 
$(3.2\pm 0.1)\times 10^{-13}$~erg~cm$^{-2}$~s$^{-1}$,
respectively, confirming the variability of both sources, particularly for SD2.

Both sources present an excess at high energy, that could be explained by a partial 
covering model for SD1 and by a gaussian line for SD2. The partial covering has been
successfully used to explain similar hard X-ray excesses in $1$H $0707-495$ 
(Grupe et al. 2004, Gallo et al. 2004a, Tanaka et al. 2004). However, it is worth 
stressing that for SD1 the statistics is not enough to reach firm conclusions. 

The large gaussian emission line used to fit the excess in SD2 can be the indication
of light bending around a Kerr black hole, like, for example, the case of PHL $1092$ 
(Gallo et al. 2004b) or reflection from ionized disc, used also for $1$H $0707-495$
(Fabian et al. 2004). Indeed, for SD2, the constant density ionized disc model
by Ross \& Fabian (1993) gives acceptable fits, even though worse with respect to the
simpler phenomenological model of the gaussian emission line. 

Dedicated observations with higher statistics are needed to assess the spectral
parameters of SD1 and SD2. Moreover, in order to understand if the characteristisc
of NLS1 are biased by the X-ray selection, it is necessary to enlarge the sample
of observed galaxies in X-ray from the optically selected sample of Williams et al.
(2002).

\begin{acknowledgements}
LF acknowledges partial financial support by the Italian Space Agency
(ASI). 

LF wishes to thank M. Cappi, P. Severgnini, and R. Williams for useful discussions.

This publication has made use of public data obtained from the High 
Energy Astrophysics Science Archive Research Centre (HEASARC), provided by 
NASA Goddard Space Flight Centre.
\end{acknowledgements}


\begin{thebibliography}{}
\bibitem[1996]{BOLLER} Boller T., Brandt W.N., Fink H., 1996, A\&A 305, 53

\bibitem[1989]{CARDELLI} Cardelli J.A, Clayton G.C., Mathis J.S., 1989, ApJ 345, 245

\bibitem[2000]{COX} Cox A.N. (editor), 2000, Allen's Astrophysical Quantities. 
IV Edition, Springer, New York

\bibitem[2003]{cutri} Cutri R.M., Skrutskie M.F., van Dyk S., et al., 2003, 
2MASS All--Sky Catalog of Point Sources. University of Massachusetts and 
Infrared Processing and Analysis Center, (IPAC/California Institute of Technology)

\bibitem[1978]{DK} Davidson M.K. \& Kinman T.D., 1978, ApJ 225, 776

\bibitem[1990]{DICKEY} Dickey J.M. \& Lockman F.J., 1990, ARAA 28, 215

\bibitem[2004]{FABIAN} Fabian A.C., Miniutti G., Gallo L., et al., 2004, MNRAS, accepted
for publication (\texttt{astro-ph/0405160})

\bibitem[2004]{GALLO1} Gallo L.C., Tanaka Y., Boller T., et al., 2004a, MNRAS, accepted
for publication (\texttt{astro-ph/0405159})

\bibitem[2004]{GALLO2} Gallo L.C., Boller T., Brandt W.N., et al., 2004b, MNRAS 352, 744

\bibitem[1989]{GOODR} Goodrich R.W., 1989, ApJ 342, 224

\bibitem[2000]{GRUPE3} Grupe D., 2000, New Astr. Rev. 44, 455

\bibitem[1998]{GRUPE1} Grupe D., Beuermann K., Thomas H.C., et al., 1998, A\&A 330, 25

\bibitem[1999]{GRUPE2} Grupe D., Beuermann K., Mannheim K., Thomas H.C., 1999, A\&A 350, 805

\bibitem[2004]{GRUPE4} Grupe D., Mathur S., Komossa S., 2004, ApJ 127, 3161

\bibitem[2003]{KIRSCH} Kirsch M., 2003. EPIC status of calibration and data analysis.
XMM-SOC-CAL-TN-0018, v. 2.1, 4 April 2003.

\bibitem[1999]{LEIGHLY1} Leighly K.M., 1999a, ApJS 125, 297

\bibitem[1999]{LEIGHLY2} Leighly K.M., 1999b, ApJS 125, 317

\bibitem[1995]{PEXRIV} Magdziarz P. \& Zdziarski A.A., 1995, MNRAS, 273, 837

\bibitem[2003]{monet2} Monet D.G., Levine S.E., Casian B., et al., 2003, AJ 125, 984

\bibitem[1985]{OP} Osterbrock D.E. \& Pogge R.W., 1985, ApJ 297, 166

\bibitem[2003]{PIER} Pier J.R., Munn J.A., Hindsley R.B., et al., 2003, AJ 125, 1559

\bibitem[2000]{POGGE} Pogge R.W., 2000, New Astr. Rev. 44, 381

\bibitem[1992]{EP} Puchnarewicz E.M., Mason K.O., C\'ordova F., et al., 1992, MNRAS 256, 589

\bibitem[1993]{RF} Ross R.R. \& Fabian A.C., 1993, MNRAS 261, 74

\bibitem[2002]{XMMABC} Snowden S., Still M., Harrus I. et al., 2002. An introduction to
XMM-Newton data analysis. Version 1.3, 26 September 2002.

\bibitem[1989]{Stephens} Stephens S., 1989, AJ 97, 10

\bibitem[2001]{STRUDER} Str\"uder L., Briel U., Dennerl K., et al., 2001, A\&A 365, L18

\bibitem[2004]{TANAKA} Tanaka Y., Boller T., Gallo L., et al., 2004, (\texttt{astro-ph/0405158})

\bibitem[2001]{TURNER} Turner M.J., Abbey A., Arnaud M., et al., 2001, A\&A 365, L27

\bibitem[1996]{WANG} Wang T., Brinkmann W., Bergeron J., 1996, A\&A 309, 81

\bibitem[2002]{WPM} Williams R.J., Pogge R.W., Mathur S., 2002, AJ 124, 3042

\end{thebibliography}
\end{document}